\documentclass[%
 aip,
 amsmath,amssymb,
 reprint,%
]{revtex4-1}
\usepackage{graphicx}
\usepackage{dcolumn}
\usepackage{bm}

\usepackage[utf8]{inputenc}
\usepackage[T1]{fontenc}
\usepackage{mathptmx}
\usepackage{etoolbox}
\usepackage{color}

\newcommand{\T}[1]{{\tt #1}}
\newcommand{\V}[1]{\mathbf{#1}}
\newcommand{\mvec}[1]{\mathbf{#1}}
\newcommand{\gvec}[1]{\boldsymbol{#1}}
\newcommand{\B}[1]{\boldsymbol{#1}}

\newcommand{\zhat}{\mbox{$\hat{\mathbf{z}}$}}

\newcommand{\figref}[1]{Fig.~\ref{#1}}   
\newcommand{\eqr}[1]{Eq.~\eqref{#1}}   
\newcommand{\secref}[1]{\S\ref{#1}}
   
\newcommand{\gke}{{\tt Gkeyll}}

\DeclareMathOperator{\sech}{sech}

\newcommand{\new}[1]{{#1}}

\makeatletter
\def\@email#1#2{%
 \endgroup
 \patchcmd{\titleblock@produce}
  {\frontmatter@RRAPformat}
  {\frontmatter@RRAPformat{\produce@RRAP{*#1\href{mailto:#2}{#2}}}\frontmatter@RRAPformat}
  {}{}
}%
\makeatother
\begin{document}

\preprint{AIP/123-QED}

\title[Electron Energization in Reconnection]{Electron Energization in Reconnection: Eulerian versus Lagrangian Perspectives}
\author{Jason M. TenBarge}
\email{tenbarge@princeton.edu}
 \affiliation{Department of Astrophysical Sciences, Princeton University, Princeton, NJ 08544, USA}
\author{James Juno}%
\affiliation{Princeton Plasma Physics Laboratory, Princeton, NJ 08540, USA}%
\author{Gregory G. Howes}
\affiliation{Department of Physics and Astronomy,University of Iowa, Iowa City IA 54224, USA}

\date{\today}

\begin{abstract}
 Particle energization due to magnetic reconnection is an important unsolved problem for myriad space and astrophysical plasmas. Electron energization in magnetic reconnection has traditionally been examined from a particle, or Lagrangian, perspective using particle-in-cell (PIC) simulations. Guiding-center analyses of ensembles of PIC particles have suggested that Fermi (curvature drift) acceleration and direct acceleration via the reconnection electric field are the primary electron energization mechanisms. However, both PIC guiding-center ensemble analyses and spacecraft observations are performed in an Eulerian \new{perspective}. For this work, we employ the continuum Vlasov-Maxwell solver within the \gke\ simulation framework to re-examine electron energization from a kinetic continuum, Eulerian, perspective. We separately examine the contribution of each drift energization component to determine the dominant electron energization mechanisms in a moderate guide-field \gke\ reconnection simulation. In the Eulerian perspective, we find that the diamagnetic and agyrotropic drifts are the primary electron energization mechanisms away from the reconnection x-point, where direct acceleration dominates.  We compare the Eulerian (Vlasov \gke) results with the wisdom gained from Lagrangian (PIC) analyses.\end{abstract}

\maketitle

\section{Introduction}\label{sec:intro}
Magnetic reconnection is a ubiquitous process in space and astrophysical plasmas, and it plays a fundamental role in transforming stored magnetic energy into particle kinetic and thermal energies. Reconnection is thought to produce high energy, non-thermal particles in a variety of systems, including gamma ray bursts \cite{Drenkhahn:2002}, solar and stellar flares \cite{Lin:2003}, and pulsar magnetospheres \cite{Michel:1994}. Non-thermal electrons are responsible for producing a significant fraction of the emitted light from high-energy astrophysical objects; therefore, understanding how these non-thermal electrons are produced is fundamentally important. Solar observational results suggest that reconnection very efficiently produces non-thermal electrons \cite{Krucker:2010,Oka:2013}. The process often associated with the acceleration of the electrons is Fermi acceleration due to the strongly curved, contracting magnetic island structures that form in the outflow of magnetic reconnection \cite{Drake:2006,Guo:2014}. These outwardly flowing magnetic structures are believed to lead to local enhancements of curvature drift acceleration. Similar structures are observed in relativistic turbulence simulations \cite{Zhdankin:2018,Comisso:2018b}, and are also sites of enhanced thermal and non-thermal electron production.

In addition to non-thermal electron production, reconnection also leads to significant thermal energization of both ions and electrons. Heating due to reconnection likely plays an important role in the heating of the solar corona and acceleration of the solar wind \cite{Parker:1983,Parker:1988,Drake:2012}. Electron heating due to small-scale reconnecting current sheets in turbulence plays a fundamental role in dissipating energy injected at large scales in myriad space and astrophysical plasmas \cite{TenBarge:2013a,Karimabadi:2013,Chasapis:2017,Shay:2018}. Recent spacecraft missions like Cluster\cite{Escoubet:1997}, THEMIS\cite{Angelopoulos:2008}, and the Magnetospheric Multiscale (MMS) mission\cite{Burch:2016b} have also identified significant electron heating (and particle acceleration) at reconnection sites in the magnetosheath \cite{Phan:2013,Burch:2016} and in the magnetotail \cite{Chen:2008,Sitnov:2019,Oka:2022,Oieroset:2023}. 

Whether thermal or non-thermal, electron energization has traditionally been examined from a particle, or Lagrangian, perspective using particle-in-cell (PIC) simulations. Guiding-center based analyses of the PIC simulations \new{have} suggested that Fermi (curvature drift) acceleration and direct acceleration via the reconnection electric field are the primary electron energization mechanisms \cite{Dahlin:2014,Dahlin:2015,Dahlin:2017,Li:2015,Li:2017}. Due to its spatially limited extent in the vicinity of the reconnection x-point or x-line, direct acceleration is generally believed to play a sub-dominant role in reconnection, except when the guide field is much larger than the reconnection field \cite{Dahlin:2016,Mccubbin:2022}. 

The guiding-center analyses are traditionally based on the work of \citet{Northrop:1963}, who semi-rigorously derives the energy evolution equation for a single electron in the non-relativistic limit, averaged over a gyration period. However, the derivation includes only the $\nabla B$ and curvature drifts, while excluding other single-particle drifts. The single-particle energization equation is then summed over local ensembles of guiding-centers to produce a fluid-like energization equation for the particle moments, i.e., $dE_e / dt = \V{j}_e \cdot \V{E}$, where $\V{j}_e$ is the electron current density, $\V{E}$ is the electric field, and $E_e$ is the total electron kinetic energy. \citet{Li:2015} effectively extend the number of drifts included in the guiding-center model by adding drifts to the electron current following derivations by \citet{Parker:1957,Blandford:2014} for the electron drift contributions to the current in a pressure gyrotropic plasma. \citet{Li:2017} adds additional drift contributions due to pressure agyrotropy. In all of these studies, guiding-center ensemble averaged analyses of the PIC results \new{indicate} curvature drift is generally the dominant guiding-center energization mechanism. Interestingly, \citet{Li:2018} performs a different separation of the drift energization mechanisms into compressional and shear (curvature drift accounting for pressure anisotropy) and find that compressional energization dominates for weak-to-moderate guide fields ($B_g \le 0.2 B_0$), while for larger guide fields, the compressional and shear terms contribute equally.

\new{However, PIC simulations follow individual particles rather than guiding centers, so particle energization terms due to the motion of the particles relative to their guiding-center rest frames are absent from existing guiding-center analyses.} Further, by performing local ensemble averages and then examining particle moments, one has moved from a Lagrangian to an Eulerian perspective. Therefore, existing kinetic studies  to identify the dominant electron energization mechanism  in reconnection miss energization terms due to being performed in the particle rather than guiding-center \new{limit}, and the studies mix Lagrangian and Eulerian perspectives. Formally, in the Eulerian perspective, one is no longer directly quantifying particle energization, rather the Eulerian perspective provides information about the particle distribution function, as well as bulk flow and thermal energization. To date, the only fully Eulerian reconnection simulations focused on identifying the primary energization mechanism have been performed using magnetohydrodyanmics (MHD). The MHD simulations examining the importance of magnetic curvature energization on bulk flow energization arrive at conflicting conclusions, finding both that curvature  dominates the energization \cite{Beresnyak:2016} and that curvature  and magnetic pressure expansion contribute approximately equally \cite{Du:2022}.

Recent phase-space diagnostics like the field-particle correlation\cite{Klein:2016,Howes:2017} (FPC)  have been developed from an Eulerian perspective to identify the signatures of particle energization due to mechanisms such as Landau \cite{Klein:2017,Chen:2019} and cyclotron \cite{Klein:2020} damping, direct acceleration in reconnection \cite{Mccubbin:2022}, and shock drift acceleration \cite{Juno:2021,Juno:2023}. However, the initial identification of a particular FPC signature requires knowledge of the active energization mechanism at a specific configuration point. Therefore, it is essential to derive an Eulerian picture of the energization to interpret velocity-space signatures provided by the FPC. Development and tabulation of these signatures will enable single-spacecraft observations to explore the nature of the electron energization in reconnection, turbulence, and shocks.

To address this dichotomy between Lagrangian and Eulerian perspectives of electron energization in reconnection, we use the continuum (Eulerian) Vlasov-Maxwell solver within \gke~to perform  2D-3V kinetic simulations of magnetic reconnection with a moderate guide field. We identify the dominant energization mechanisms locally and globally in an Eulerian simulation by employing an Eulerian-based analysis, including a rigorously complete description of all fluid drifts: diamagnetic, curvature, polarization, and agyrotropic. The results herein are straightforwardly generalizable to 3D for use in other simulation or spacecraft data analysis.

In \secref{sec:model}, we derive and discuss the physical significance of the guiding-center and Eulerian energization equations, including the presentation of alternative formulations of the Eulerian energization. We describe the continuum kinetic simulation code, \gke, and initial conditions in \secref{sec:simulations}. In \secref{sec:results}, we present the results of the \gke~simulations. Finally, in \secref{sec:conclusions}, we discuss the implications and significance of the results.

\section{Energization Models}\label{sec:model}
\subsection{Guiding-Center Energization}
To examine the components of electron energization, we begin by following the approach established by \citet{Dahlin:2014,Dahlin:2015,Dahlin:2017} and \citet{Northrop:1963} to construct the energy evolution equation for a single electron in the the non-relativistic, guiding-center limit averaged over a gyration period,
\begin{equation}\label{eq:single_part}
    \frac{d \epsilon}{dt} = q_e v_\parallel \V{b} \cdot \V{E} + \mu \frac{\partial B}{\partial t} + q_e \left(\V{v}_g + \V{v}_c\right) \cdot \V{E}.
\end{equation}
Here, $B = |\V{B}|$, $\V{b} = \V{B} / B$, $v_\parallel = \V{v} \cdot \V{b}$, $\V{v}_\perp = \V{v} - v_\parallel \V{b}$, $q_e$ and $m_e$ are the electron charge and mass, $\mu = m_e v_\perp^2 / 2 B$ is the magnetic moment, $\V{v}_c$ and $\V{v}_g$ are the curvature and grad-B drifts,
\begin{equation}\label{eq:curvDrift}
    \V{v}_c = \frac{v_\parallel^2}{\Omega_{c_e}} \V{b} \times \B{\kappa},
\end{equation}
\begin{equation}\label{eq:gradDrift}
    \V{v}_g = \frac{v_\perp^2}{2 \Omega_{c_e}} \V{b} \times \frac{\B{\nabla} B}{B},
\end{equation}
$\B{\kappa} = \V{b} \cdot \B{\nabla} \V{b}$ is the magnetic curvature, and $\Omega_{ce} = q_e B / m_e$ is the electron cyclotron frequency. Finally, $\epsilon = m v_\parallel^2 / 2 + m u_E^2 /2 + \mu B$, where $\V{u}_E = \V{E} \times \V{B} / B^2$ is the E cross B drift. To lowest order, the first two terms in $\epsilon$ correspond to the guiding-center kinetic energy, while $\mu B$ is related to the average energy of rotation about the guiding center. 

Summing \eqr{eq:single_part} over all guiding centers in a local region, i.e., an ensemble of guiding centers, provides an evolution equation of the total electron kinetic energy,
\begin{eqnarray}\label{eq:total_kinetic}
    \frac{d E_e}{dt}  &=& j_\parallel E_\parallel + \frac{p_\perp}{B} \left(\frac{\partial B}{\partial t} + \V{u}_E \cdot \B{\nabla} B\right) +  (p_\parallel + m_e n u_\parallel^2) \V{u}_E \cdot \B{\kappa} \nonumber \\
    &=& W_0 =  W_{par} + W_{beta-gradB} + W_{curv0}.
\end{eqnarray}
The  first term on the right hand side (RHS) of \eqr{eq:total_kinetic} corresponds to direct electron energization by the parallel electric field. The second term, betatron plus $\nabla B$ acceleration, provides perpendicular  energization and is responsible for conserving the magnetic moment---we will simply refer to this term as betatron-$\nabla B$ acceleration. The final curvature related term gives rise to the first-order Fermi acceleration mechanism described in \citet{Drake:2006}. 

Before describing the drifts in an Eulerian \new{perspective}, we note that this guiding-center approach is somewhat ambiguous because only the curvature and $\nabla B$ drifts have been included. \citet{Li:2015,Li:2017} ameliorate this shortcoming by including additional drifts, some of which they find to be particularly important in the weak or zero guide field limits. However, regardless of the number of drifts included, \eqr{eq:total_kinetic} only accounts for the kinetic energy of the guiding centers plus the particle energy averaged over the gyroperiod.  


\subsection{Eulerian Energization}
 While the guiding-center approach has been applied successfully to PIC simulations, it is not directly applicable to continuum kinetic or fluid simulations, which are performed in an Eulerian \new{framework} and describe distributions of particles rather than ensembles. To determine the drift decomposition for Eulerian simulations, one must turn to the momentum equation for species $s$ to find the bulk drifts
\begin{equation}
    \frac{d\mvec{u}_s}{dt} + \frac{1}{m_s n_s}\nabla\cdot\mvec{P}_s = \frac{q_s}{m_s}(\mvec{E} + \mvec{u}_s\times\mvec{B}), \label{eq:momentumEq}
\end{equation}
where
\begin{equation}
    \frac{d}{dt} \equiv \frac{\partial}{\partial t} + \mvec{u}_s\cdot\nabla,
\end{equation}
and $\mvec{P}_s$ is the pressure tensor,
\begin{equation}
    \mvec{P}_s  = m_s \int (\mvec{v} - \mvec{u}_s) (\mvec{v} - \mvec{u}_s) f_s d\mvec{v}.
\end{equation}
If the plasma is magnetized, it is natural to split the pressure tensor as 
\begin{equation}
    \mvec{P}_s = \mvec{P}^G_s + \gvec{\Pi}_s^a,
\end{equation}
where
\begin{equation}
    \mvec{P}^G_s = \mvec{I} p_{s,\perp} + \mvec{b}\mvec{b} (p_{s,\parallel}-p_{s,\perp}) 
\end{equation}
is the gyrotropic or Chew-Goldberger-Low (CGL) pressure tensor \cite{Chew:1956} and $\gvec{\Pi}_s^a$ is the remaining, agyrotropic portion of the pressure tensor. Finally, to compute the drifts, one must cross \eqr{eq:momentumEq} with $\mvec{B}$. After manipulation (see Appendix C of \citet{Juno:2021} for details), 
\begin{eqnarray}
    \mvec{u}_{s,\perp} &=&
    \frac{\mvec{E}\times\mvec{B}}{|\mvec{B}|^2}
    + \frac{\mvec{B} \times \nabla p_{s,\perp}}{q_s n_s |\mvec{B}|^2}  
    +  \Delta p_s \frac{\mvec{B} \times \nabla_\parallel \mvec{b}}{q_s n_s |\mvec{B}|^2}  
    + \frac{\mvec{B} \times \nabla\cdot\gvec{\Pi}_s^a}{q_s n_s |\mvec{B}|^2}
    +  \nonumber \\
    & & \frac{m_s}{q_s |\mvec{B}|^2}\mvec{B} \times \frac{d\mvec{u}_s}{dt},
  \label{eq:uperp}
\end{eqnarray}
where $\Delta p_s = (p_{s,\parallel}-p_{s,\perp})$ is the pressure anisotropy. In \citet{Juno:2021}, the RHS terms of \eqr{eq:uperp} are referred to as the $\mvec{E} \times \mvec{B}$, diamagnetic, curvature, agyrotropic, and polarization drifts, respectively. We note that while there is some ambiguity in this nomenclature, \eqr{eq:uperp} is a complete description of all perpendicular bulk motion in the plasma.

We can further construct the particle energization by dotting $\mvec{u}_s$ with $q_s n_s \mvec{E}$, where $\mvec{u}_{s,\perp}$ is defined by \eqr{eq:uperp}, yielding
\begin{eqnarray}\label{eq:fluidEnergization}
    &\mvec{j}_s \cdot \mvec{E}& = W_1 = W_\parallel + W_{diam} + W_{curv1} + W_{agyro} + W_{pol} :=\\ 
    &j_{\parallel,s} E_\parallel &
    + \mvec{u}_E \cdot \nabla p_{s,\perp}
    + \Delta p_s \mvec{u}_E \cdot \B{\kappa}
    + \mvec{u}_E \cdot (\nabla \cdot\gvec{\Pi}_s^a)
    + m_s n_s \mvec{u}_E \cdot \frac{d\mvec{u}_s}{dt}. \nonumber
\end{eqnarray}
Following the bulk drift labelling scheme, the RHS terms correspond to energization via parallel, diamagnetic, curvature, agyrotropic, and polarization drifts. 

By comparing to the guiding-center energization equation, \eqr{eq:total_kinetic}, to \eqr{eq:fluidEnergization} one significant difference is obvious: the bulk curvature drift, $W_{curv1}$, depends on the pressure anisotropy rather than simply $p_\parallel$. This dependence of the bulk curvature drift on the pressure anisotropy may seem strange; however, it arises due to the additional motion of the particles \new{relative to their} guiding-center rest frames that must be added to the guiding-center drifts to transform to the Eulerian frame (see Ch. 7 of \citet{Goldston:1995} for a full derivation). Despite this physical difference in \new{reference} frames,  we can re-organize the terms in \eqr{eq:fluidEnergization} to more closely resemble the guiding-center derivation:
\begin{eqnarray}\label{eq:fluidEnergization2}
    &\mvec{j}_s \cdot \mvec{E}& = W_2 = W_\parallel + W_{beta} + W_{curv2} + W_{agyro} + W_{pol} := \nonumber \\ 
    &j_{\parallel,s} E_\parallel& 
    + (\mvec{u}_E \cdot \nabla p_{s,\perp}
    - p_{s,\perp} \mvec{u}_E \cdot \B{\kappa})
    + p_{s,\parallel} \mvec{u}_E \cdot \B{\kappa}
    + \\
    &\mvec{u}_E \cdot (\nabla \cdot\gvec{\Pi}_s^a)&
    + m_s n_s \mvec{u}_E \cdot \frac{d\mvec{u}_s}{dt}. \nonumber
\end{eqnarray}
In this formulation, the curvature term, $W_{curv2}$, contains only the parallel pressure, and the betatron-like term, $W_{beta}$ now contains \new{all of the} $p_\perp$ dependence, including curvature-related components. To make the comparison more direct, we can also further manipulate the guiding-center drift energization equation (see \eqr{eq:total_kinetic2} in Appendix~\ref{sec:appendix}). 

Motivated by the alternative guiding-center drift energization equation,  \eqr{eq:total_kinetic2}, we can perform a final, alternative re-organization of the fluid-drift energization equation, 
\begin{eqnarray}\label{eq:fluidEnergization3}
    &\mvec{j}_s \cdot \mvec{E}& = W_3 = W_\parallel + W_{diam} +  W_{beta-curv} + W_{curv2} + \nonumber\\
    &W_{agyro}& + W_{pol} :=     j_{\parallel,s} E_\parallel 
    + \mvec{u}_E \cdot \nabla p_{s,\perp}
    - p_{s,\perp} \mvec{u}_E \cdot \B{\kappa}
    + \nonumber \\ 
    &p_{s,\parallel} \mvec{u}_E \cdot \B{\kappa}&
    + \mvec{u}_E \cdot (\nabla \cdot\gvec{\Pi}_s^a)
    + m_s n_s \mvec{u}_E \cdot \frac{d\mvec{u}_s}{dt}. 
\end{eqnarray}
This formulation separates the two curvature-related drift energization terms, which will make later comparison of the drifts more transparent.

\begin{figure}
    \includegraphics[width=\linewidth]{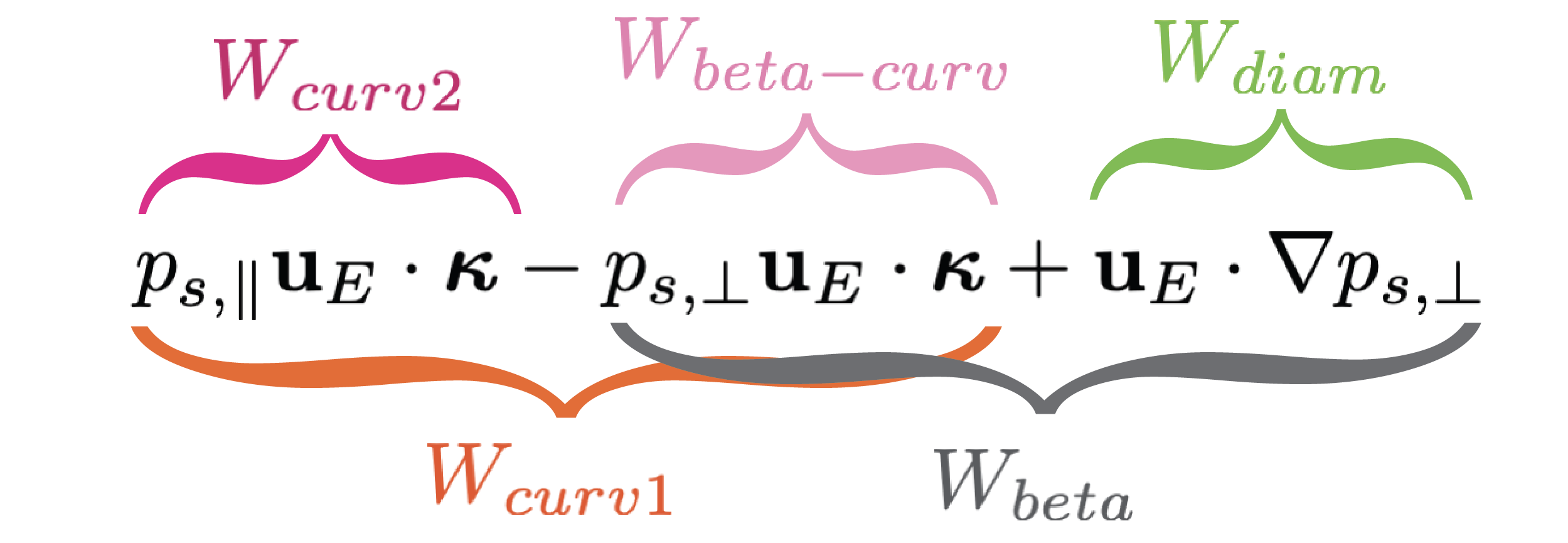}
    \caption{Illustration of the different groupings of the terms in Eqs.~ \ref{eq:fluidEnergization}, \ref{eq:fluidEnergization2}, and \ref{eq:fluidEnergization3}, where $W_{curv2}=W_{curv0}-m_sn_su_{s,\parallel} \mathbf{u}_E \cdot \boldsymbol{\kappa}$ is the Lagrangian curvature drift term minus the curvature term that appears in the polarization drift in \eqr{eq:polarization_curv}, where the color-coding corresponds to the legend of \figref{fig:IntegratedBg}.} 
    \label{fig:Curv_terms}
\end{figure}

To aid in following the three different groupings of the curvature, betatron, and diamagnetic drift terms, \figref{fig:Curv_terms} provides a visual illustration of the groupings. The color-coding corresponds to that used in \figref{fig:IntegratedBg}.

Finally, we note that the $m n u_\parallel^2$ curvature term appearing in \eqr{eq:total_kinetic} is contained in the Eulerian polarization drift term, since
\begin{eqnarray}\label{eq:polarization_curv}
    m_s n_s \mvec{u}_E \cdot (\mvec{u_{s,\parallel}} \cdot \nabla \mvec{u_{s,\parallel}}) &=& m_s n_s \mvec{u}_E \cdot (u_{s,\parallel}^2 \B{\kappa} + \mvec{u}_{s,\parallel} \nabla_\parallel u_{s,\parallel}) \nonumber \\ 
    &=& m_s n_s u_{s,\parallel}^2 \mvec{u}_E \cdot \B{\kappa},
\end{eqnarray}
where $\mvec{u_{s,\parallel}} = u_{s,\parallel} \mvec{b} \equiv (\mvec{u}_s \cdot \mvec{b}) \mvec{b}$. We find below that $W_{pol}$ contributes little to the electron energization. Since $W_{pol}$ is small, this term is itself either small or mostly cancelled by other terms appearing in $W_{pol}$. Therefore, we do not separate it when discussing Eulerian energization.

\section{Simulation Setup}\label{sec:simulations}
To study electron energization in magnetic reconnection, we employ the fully kinetic continuum Vlasov-Maxwell solver portion of the \gke\ framework \cite{Juno:2018,Hakim:2020b}. \gke\ directly evolves the full Vlasov-Maxwell system in up to 3D-3V, three configuration and three velocity space dimensions, for an arbitrary number of species using the discontinuous Galerkin discretization scheme in all phase space dimensions. 

To initialize magnetic reconnection, we follow the basic initial conditions of the Geospace Environment Modeling (GEM) reconnection challenge \cite{Birn:2001}. Therefore, we initialize a Harris sheet with the following magnetic field and density profiles:
\begin{equation}
    B_x(y) = B_0 \tanh{(y/\lambda)},
\end{equation}
\begin{equation}
    n(y) = n_0 \sech^2{(y/\lambda)} + n_\infty,
\end{equation}
with uniform ion and electron temperatures, $T_i$ and $T_e$, satisfying $T_i / T_e = 5$. To maintain electron magnetization throughout the layer \cite{Swisdak:2005}, we add a uniform, moderate guide field, $\mvec{B}_g = 0.1 B_0 \hat{\mvec{z}}$, to the above magnetic field profile. Pressure balance requires that $n_0 (T_i + T_e) = B_0^2 / 2 \mu_0$, which implies $\beta_i = 2 \mu_0 n_0 T_i / B_0^2 = 5/6$. We choose $n_0/n_\infty = 5$, a reduced mass ratio, $m_i / m_e = 25$, and speed of light, $v_{A_e} / c = 0.25$, where $v_{A_e} = B_0 / \sqrt{ \mu_0 m_e n_e}$ is the in-plane, upstream, electron Alfv\'en speed. The layer thickness, $\lambda$, is taken to be $\lambda = 0.5 d_i$, where $d_i = v_{A_i} / \Omega_{ci}$ is the ion inertial length.  The simulation domain is chosen to be $-L_x/2 \le x \le L_x/2$, $-L_y/2 \le y \le L_y/2$, $-6v_{th_s} \le v_s \le 6 v_{th_s}$, where $L_x = 2 L_y  = 8 \pi d_i$, $v_{th_s} = \sqrt{2 T_s / m_s}$, and the same velocity space extents are used in each species' three velocity dimensions, except we extend the velocity space extents of the electrons in the guide field direction to $-8v_{th_e} \le v_{z,e} \le 8 v_{th_e}$ to resolve additional particle acceleration. Periodic boundary conditions are employed in the $x$ direction and conducting (reflecting) walls for the fields (particles) are used in the $y$ direction, while we employ zero-flux boundary conditions in velocity space. $(n_x,n_y) = (112,56)$  configuration space cells and $n_{v}^3 = 24^3$ velocity space cells are used for the protons and electrons with piecewise quadratic Serendipity polynomial basis functions \cite{Arnold:2011} to discretize the fields and particle distribution functions. We also employ a conservative Lenard-Bernstein collision operator \cite{Lenard:1958,Hakim:2020} with electron-electron collision frequency $\nu_e = 0.0004 \Omega_{c_e}$ and ion-ion collision frequency $\nu_i = \nu_e / \sqrt{m_i / m_e}$. The collision frequencies are chosen to satisfy the weakly collisional limit while being sufficient to regularize velocity space and maintain numerical stability.

To initiate reconnection, an initial magnetic flux function,
\begin{equation}
    \psi(x,y) = \psi_0 \cos{(2 \pi x/ L_x)} \cos{( \pi y/L_y)},
\end{equation}
is specified with $\psi_0 = 0.1 B_0 d_i$, where the perturbed magnetic field resulting from this flux is $\V{B}_\psi = \zhat \times \B{\nabla} \psi$. To break the symmetry inherent to these initial conditions in a noise-free simulation, the first twenty $x$ and $y$ magnetic field modes are randomly perturbed with a root-mean-square amplitude $B^{RMS}_{pert} = 0.01 B_0$.

\section{Simulation Results}\label{sec:results}
\subsection{Overall Evolution}

\begin{figure}
    \includegraphics[width=\linewidth]{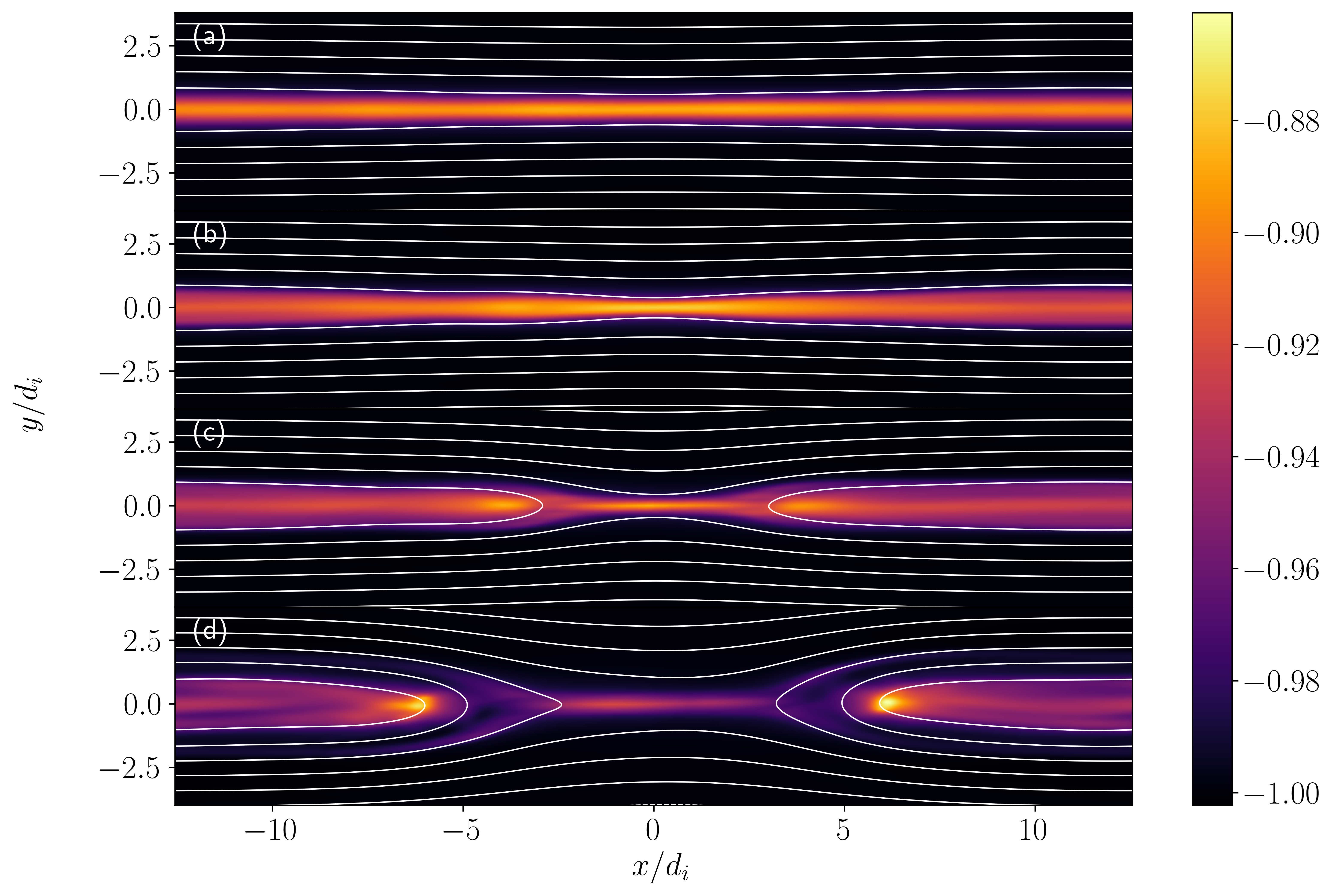}
    \caption{In panels (a)-(d) are plotted the evolution of $j_z$ in color and contours of the out-of-plane vector potential, $A_z$. Panels (a)-(d) present the evolution of the current layer at $t \Omega_{ci} =4,9,14,$ and $19$ respectively.}
    \label{fig:Jz}
\end{figure}

\begin{figure}
    \includegraphics[width=\linewidth]{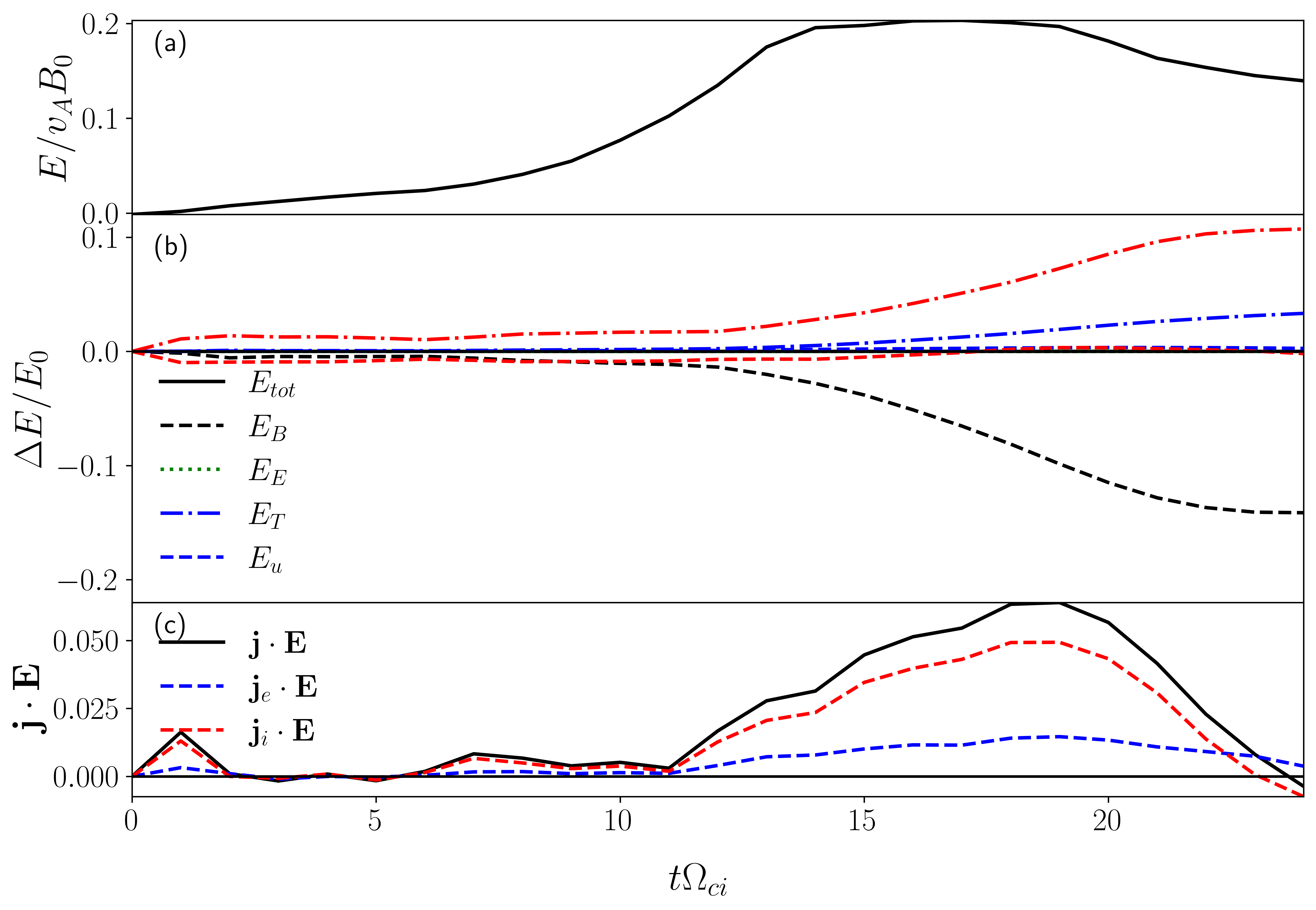}
    \caption{In panel (a) is plotted the evolution of the reconnection rate. In panel (b) are the components of the  spatially integrated and normalized change in energy, where  $E_B = B^2 / 2 \mu_0$ is the magnetic energy, $E_E = \epsilon_0 E^2 / 2$ is the electric field energy, $E_u = m n u^2 / 2$ is the flow energy, $E_T = \rm{Tr}(\V{P}) / 2$ is the thermal energy, $E_{tot} = E_B + E_E + \sum_s E_{Ts} + E_{us}$, $E_0 = E_{tot}(t=0)$, and $\Delta E_*(t) = E_*(t) - E_*(t=0)$. In panel (c) is plotted the $\mvec{j}\cdot\mvec{E}$ work. Red (blue) lines correspond to ion (electron) quantities in panels (b) and (c).}
    \label{fig:SummaryBg}
\end{figure}

 In \figref{fig:Jz}, we present the evolution of the out-of-plane current density, $j_z$, in color along with contours of the out-of-plane vector potential, $A_z$, at $t \Omega_{ci} =4,9,14,$ and $19$ in panels (a)-(d), respectively. In \figref{fig:SummaryBg}, we present the reconnection rate (a), integrated and normalized change in energy components (b), and the $\mvec{j}\cdot\mvec{E}$ work (c). The components of the  spatially integrated and normalized change in energy in panel (b) are defined as:  $E_B = B^2 / 2 \mu_0$ is the magnetic energy, $E_E = \epsilon_0 E^2 / 2$ is the electric field energy, $E_u = m n u^2 / 2$ is the flow energy, $E_T = \rm{Tr}(\V{P}) / 2$ is the thermal energy, $E_{tot} = E_B + E_E + \sum_s (E_{Ts} + E_{us})$, $E_0 = E_{tot}(t=0)$, and $\Delta E_*(t) = E_*(t) - E_*(t=0)$. In panels (b) and (c), red (blue) lines correspond to ion (electron) quantities. From these panels, we can see that the electron work peaks at $t\Omega_{Ci} \simeq 19 $, shortly after the peak reconnection rate. We will focus on this time for our local analyses of electron energization.

\subsection{Electron Energization}
\begin{figure*}
    \includegraphics[width=\linewidth]{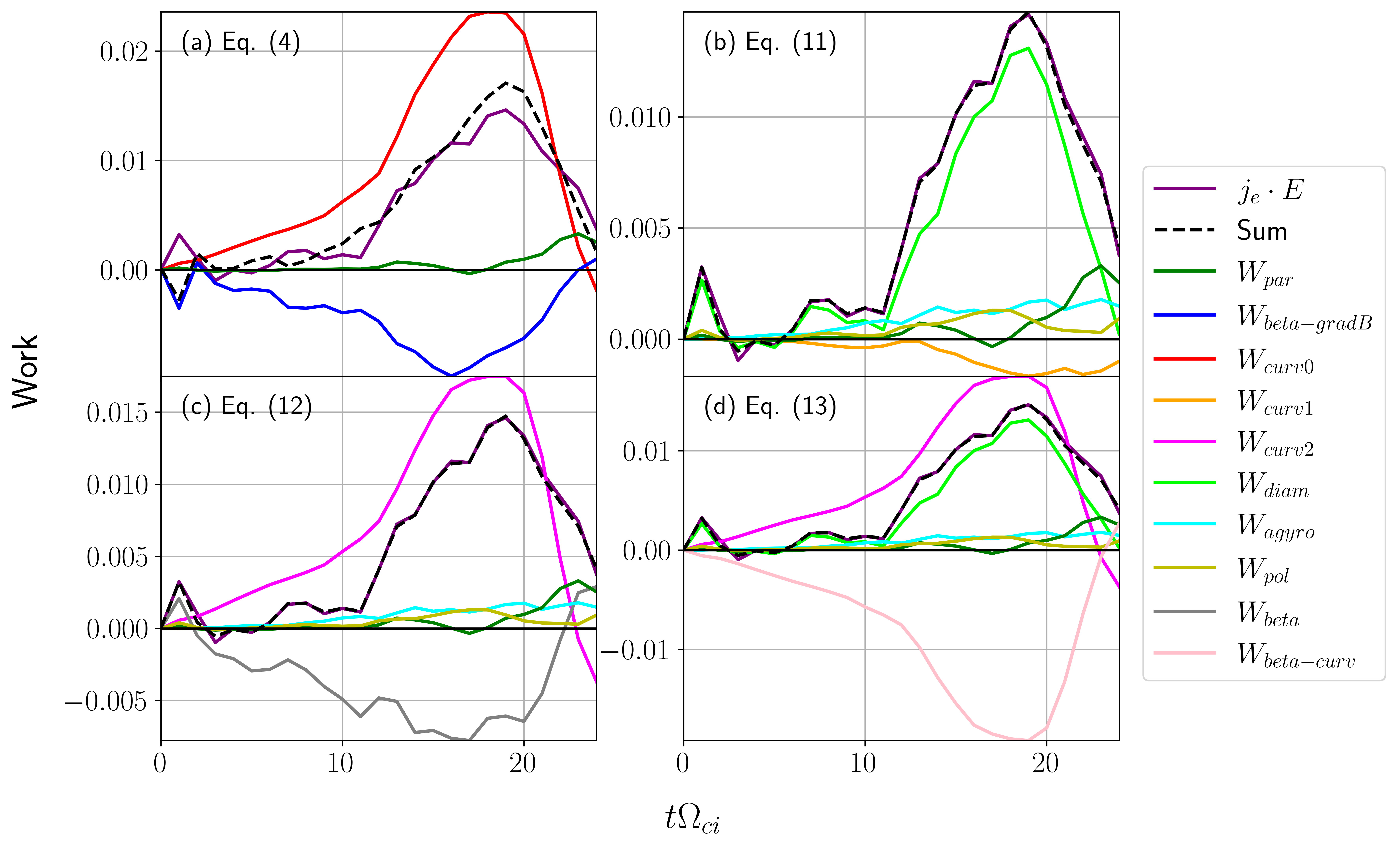}
    \caption{In each panel are presented the domain integrated electron drift energization terms appearing in different forms of the energization equation. In panel (a) is plotted the guiding center drift energization terms appearing in \eqr{eq:total_kinetic}. In panel (b)   is plotted the drift energization terms appearing in \eqr{eq:fluidEnergization}.  In panel (c)  is plotted the drift energization terms appearing in \eqr{eq:fluidEnergization2}. In panel (d), we separate the diamagnetic and curvature-like betatron, $W_{beta-curv} = - p_{s,\perp} \mvec{u}_E \cdot \B{\kappa}$, terms, as in \eqr{eq:fluidEnergization3}. The sum curve (black dashed) in each panel refers to the sum of the terms appearing in that panel.} 
    \label{fig:IntegratedBg}
\end{figure*}

 Following \cite{Dahlin:2014,Li:2015,Li:2017}, we \new{integrate over the entire domain to} examine the drift energization terms \new{and their sums} in Eqs. \ref{eq:total_kinetic}, \ref{eq:fluidEnergization}, and \ref{eq:fluidEnergization2}, which are plotted in \figref{fig:IntegratedBg}. The guiding-center energization terms (\eqr{eq:total_kinetic}) plotted in panel (a) display the expected results based on PIC simulations: the curvature drift (red) is large, positive, and the primary contributor to the sum (dashed black); the $W_{beta-gradB}$ (blue) is negative due to the conversion of magnetic energy into particle energy; and $W_{par}$ (green) is positive but small. However, the sum and the directly computed $\mvec{j}_e \cdot \mvec{E}$ (magenta) \new{track each other well but display quantitative differences of up to 20\%}, suggesting that non-negligible energization terms may be missing. 
 
 Moving to the Eulerian energization formulations in Eqs. \ref{eq:fluidEnergization} and \ref{eq:fluidEnergization2} in panels (b) and (c) respectively, we first note that the sum (black dashed) and $\mvec{j}_e \cdot \mvec{E}$ (purple) agree very well at all times, indicating that all significant energization is \new{included}, as expected for a rigorous accounting of all bulk flow motion. Next, we note that the diamagnetic drift (lime), $W_{diam}$, is the primary contributor to the domain integrated work in panel (b). For the base Eulerian energization formulation, \eqr{eq:fluidEnergization}, the curvature drift (orange), $W_{curv1}$, is at all times negative in panel (b). The agyrotropic (cyan) and polarization (yellow) drifts as well as $W_{par}$ contribute little to the total. Mirroring the guiding-center formulation using  \eqr{eq:fluidEnergization2} and splitting the curvature drift, $W_{curv1}$, into parallel curvature drift, $W_{curv2}$, (magenta) and betatron acceleration, $W_{beta}$, (grey) in panel (c), we now see that this form of the curvature drift is large and positive, dominating all other drifts, except at late times. In panel (d), we repeat the decomposition found in panel (b), except we separate the diamagnetic ($W_{diam} = \mvec{u}_E \cdot \nabla p_{s,\perp}$) and curvature-like betatron (pink), $W_{beta-curv} = - p_{s,\perp} \mvec{u}_E \cdot \B{\kappa}$, terms (see \figref{fig:Curv_terms} for a visual illustration of the groupings of curvature, betatron, and diamagnetic drifts). Here, we see clearly that the curvature-like betatron term (pink), $W_{beta-curv}$, and $W_{curv2}$ (magenta) have similar magnitudes and opposite signs, with $W_{beta-curv}$ being larger for most of the simulation, leading to $W_{curv1} = W_{curv2} + W_{beta-curv}$ being small and negative in the panel (b).

To make the comparison between drift models and their terms more quantitative, Table \ref{tab:integrated} presents the same domain integrated values as \figref{fig:IntegratedBg} but focusing on $t\Omega_{ci} = 19$. Each term of the different energization formulations is presented separately, normalized to $\V{j}_e\cdot\V{E}$, and the sum column entries refer to the sum of all contributing terms in that row. Here, we can see more clearly that the guiding-center formulation overestimates the energization by $\sim 17$\%, while the fluid energization approaches are equivalent and accurate to within $1$\%. We can also see that the diamagnetic drift contributes $\sim 90$\% of the overall energization. Finally, using \eqr{eq:fluidEnergization3}, it is clear that $W_{curv2}$ and $W_{beta-curv}$ almost cancel each other, leaving a remainder of $W_{curv1} \simeq -0.11$, an overall loss of energy due to the fluid curvature drift.

\begin{table*}
\caption{\label{tab:integrated}Domain integrated values for each term in the presented drift energization decompositions at $t\Omega_{ci} = 19$. Each term is normalized to $\V{j}_e\cdot\V{E}$, and the sum column entries refer to the sum of all contributing terms in that row.}
\begin{ruledtabular}
\begin{tabular}{ccccccccccccc}
 Equation&$\V{j}_e\cdot\V{E}$&$W_{par}$&$W_{beta-gradB}$&$W_{curv0}$&$W_{diam}$&$W_{curv1}$&$W_{agyro}$&$W_{pol}$&$W_{curv2}$&$W_{beta}$&$W_{beta-curv}$&Sum\\ \hline

$W_0$ [\eqr{eq:total_kinetic}] & 1 & 0.0484 & -0.4866 & 1.6062 & - & - & - & - & - & - & - & 1.1681\\ 

$W_1$ [\eqr{eq:fluidEnergization}] & 1 & 0.0484 & - & - & 0.8959 & -0.1144 & 0.1136 & 0.0644 & - & - & -& 1.0079\\

$W_2$ [\eqr{eq:fluidEnergization2}] & 1 & 0.0484 & - & - & - & - & 0.1136 & 0.0644 & 1.1971 & -0.4157 & - & 1.0079\\

$W_3$ [\eqr{eq:fluidEnergization3}] & 1 & 0.0484 & - & - & 0.8959 & - & 0.1136 & 0.0644 & 1.1971 & - & -1.3116 & 1.0079\\

\end{tabular}
\end{ruledtabular}
\end{table*}

\begin{figure*}
    \centering
    \includegraphics[width=\linewidth]{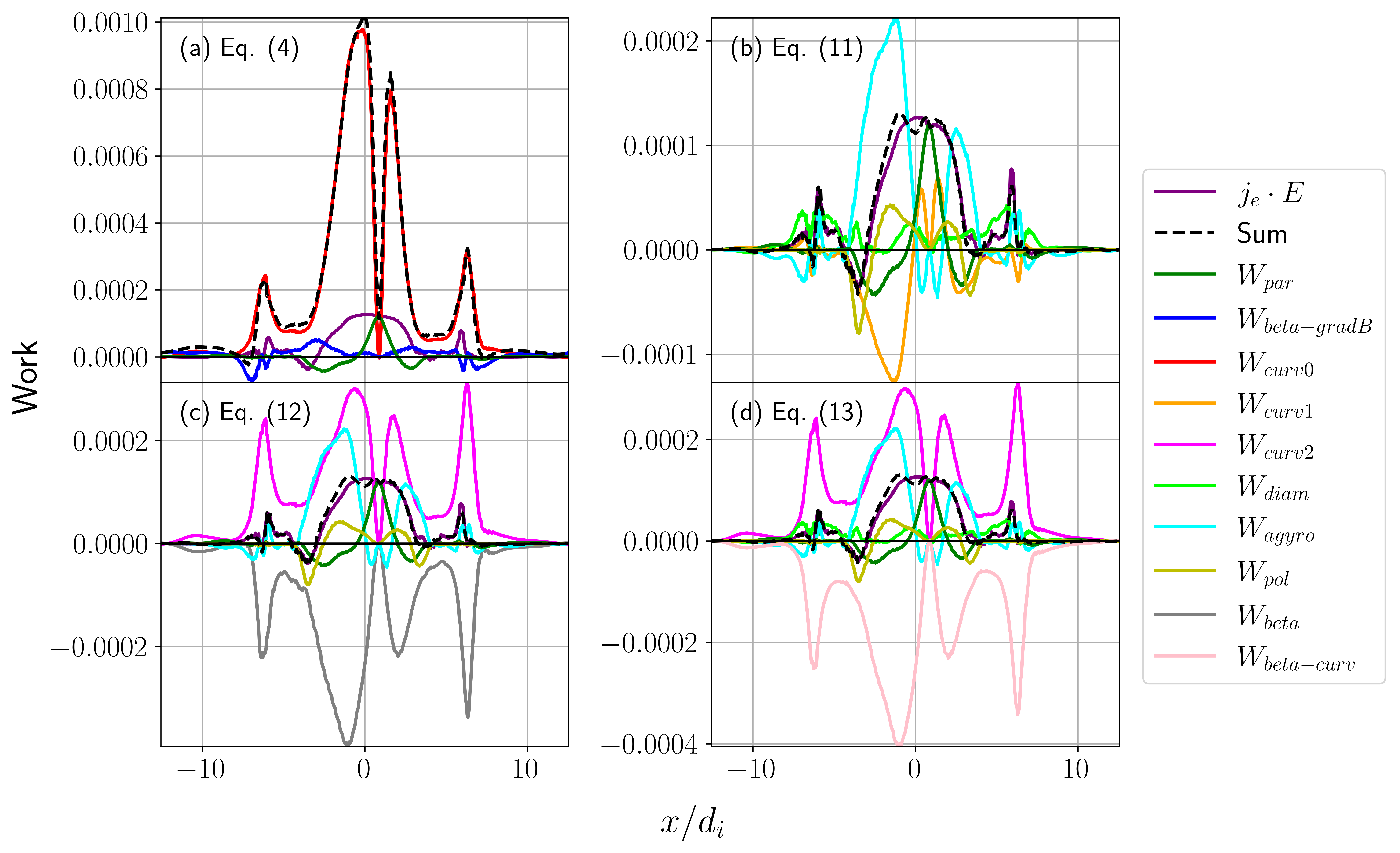}
  
    \caption{In each panel are presented cuts of the electron drift energization terms along the mid-plane, $y=0$, at $t\Omega_{ci} = 19$. In panel (a) is plotted the guiding center drift energization terms appearing in \eqr{eq:total_kinetic}. In panel (b) is plotted the energization terms appearing in \eqr{eq:fluidEnergization}. In panel (c) is plotted the energization terms appearing in \eqr{eq:fluidEnergization2}. In panel (d), we separate the diamagnetic and curvature-like betatron, $W_{beta-curv} = - p_{s,\perp} \mvec{u}_E \cdot \B{\kappa}$, terms, as in \eqr{eq:fluidEnergization3}.  The sum curve (black dashed) in each panel refers to the sum of the terms appearing in that panel.} 
    \label{fig:CutsBg}
\end{figure*}

Next, we look at cuts of the energization terms along the mid-plane at $y=0$ at the time of maximum energization ($t \Omega_{ci} = 19$). In  \figref{fig:CutsBg}, we present these cuts using the same four drift formulations presented in \figref{fig:IntegratedBg}. First, at the x-point, $x \simeq 1 d_i$, all of the energization is due to the parallel acceleration term (green), $W_{par}$. Using the guiding center formulation \eqr{eq:total_kinetic} (a), we see that away from the x-point, the curvature drift term is the largest contributor; however, we also see that the sum (dashed black) does not well agree with the directly computed $\mvec{j}_e \cdot \mvec{E}$ at any point along the cut. Using the Eulerian description \eqr{eq:fluidEnergization} (b), we see that the largest positive contributor is the agyrotropic drift (cyan), and the curvature drift (orange) is generally large and negative at the same position in $x$. Using the alternative Eulerian description \eqr{eq:fluidEnergization2} (c), again the curvature drift (magenta) is generally large and positive but is mostly balanced by an even larger, negative contribution from the betatron term (grey).  In panel (d), we again repeat the decomposition found in panel (b), except we separate the diamagnetic (lime) and curvature-like betatron terms (pink). In all cases, we note that at most points away from the x-point, several drifts contribute simultaneously with comparable amplitudes, with the agyrotropic drift generally dominating and the curvature drift ($W_{curv1}$) being generally negative.

\begin{figure*}
    \centering
    \includegraphics[width=\linewidth]{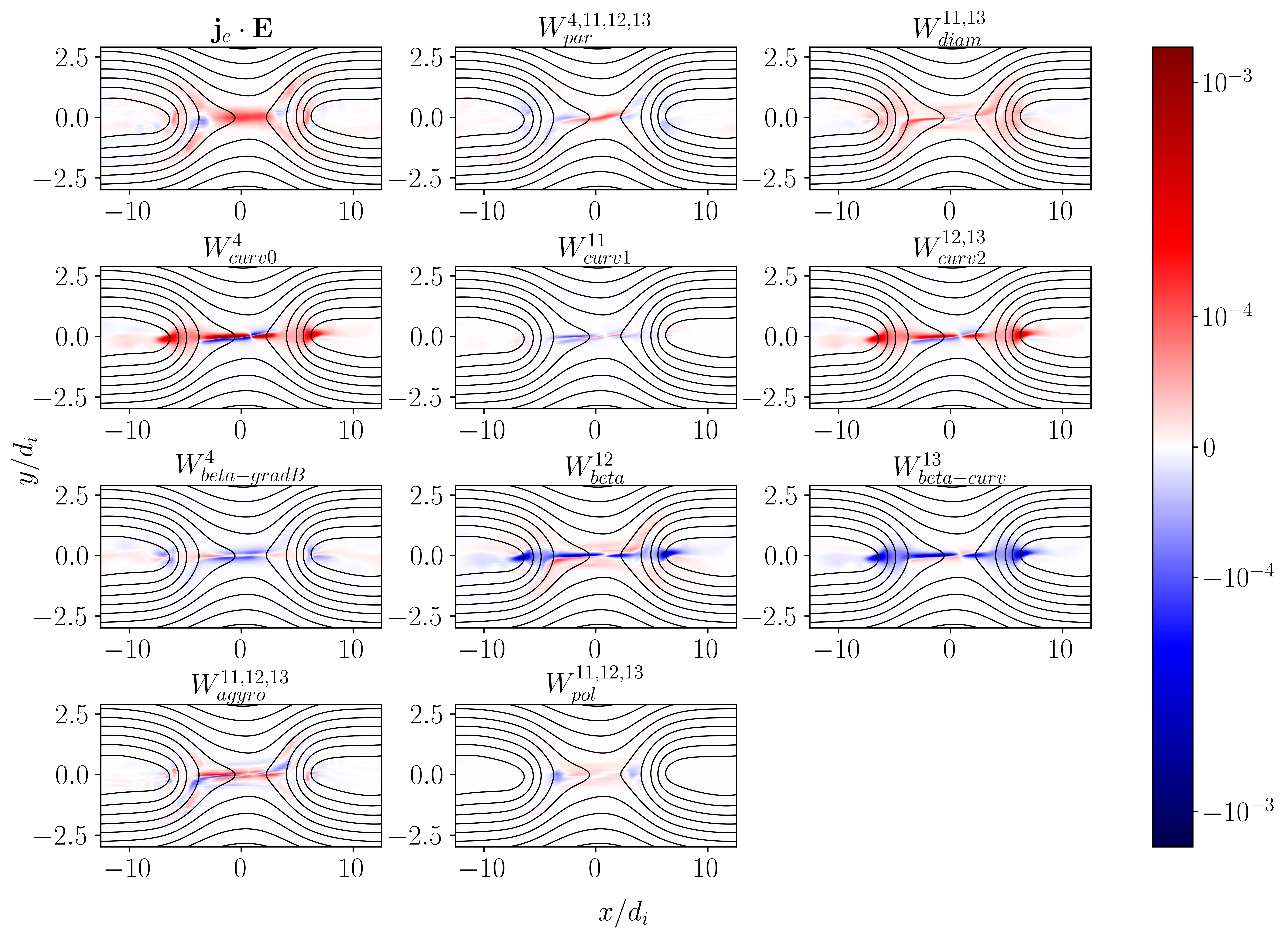}
    
    \caption{In each labelled panel, we present contour plots of the various terms contributing to electron drift energization in color with black contours of the out-of-plane vector potential, $A_z$, at $t\Omega_{ci} = 19$. Superscript values indicate the drift energization equation in which the term appears. Due to the wide range of values, we employ a mixed log-linear color bar which is logarithmic for values $|W| \geq 1 \times 10^{-4}$ and linear for values $|W| < 1 \times 10^{-4}$.} 
    \label{fig:SpatialBg}
\end{figure*}

\begin{figure}
    \includegraphics[width=\linewidth]{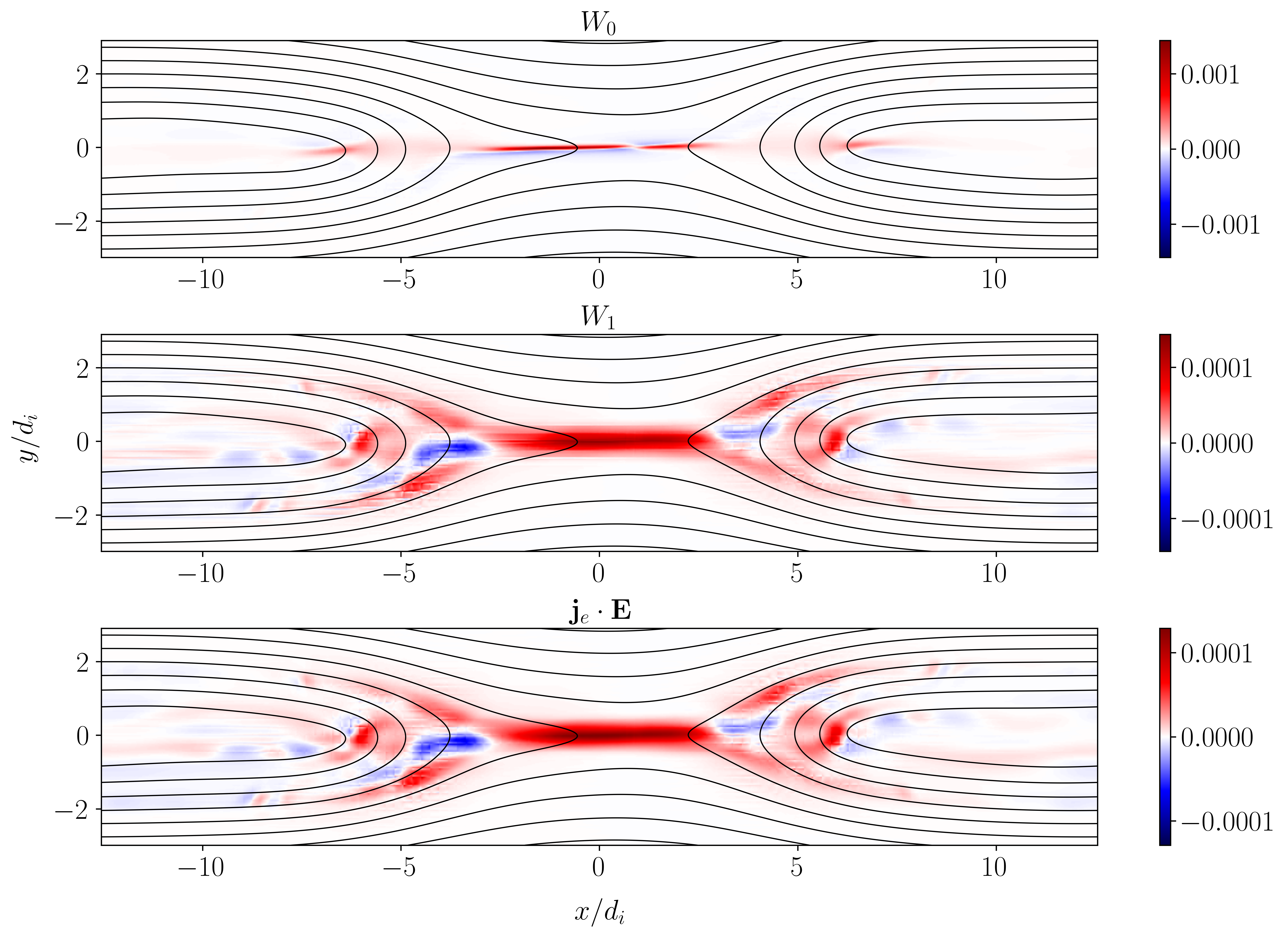}
    \caption{In each labelled panel, we present contour plots of the total electron drift energization in color with black contours of the out-of-plane vector potential, $A_z$, at $t\Omega_{ci} = 19$. In the top and middle panels, we present the guiding center sum, $W_0$, and the Eulerian sum, $W_1$. For comparison, in the bottom panel is plotted $\V{j}_e\cdot\V{E}$. }
    \label{fig:Sum}
\end{figure}

Finally, we examine the full spatial distribution of $W$ and its components at $t \Omega_{ci} = 19$ in \figref{fig:SpatialBg} with black contours of the out-of-plane vector potential, $A_z$, overplotted for reference. Superscript values indicate the drift energization equation in which the term appears. Due to the wide range of values, we employ a mixed log-linear color bar which is logarithmic for values $|W| \geq 1 \times 10^{-4}$ and linear for values $|W| < 1 \times 10^{-4}$. From these panels, we can see: i) $W_\parallel$ is large and positive in the vicinity of the x-point; ii) $W_{diam}$ is mostly positive and is the most space-filling term; iii) $W_{agyro}$ is large and positive in the entire diffusion region; iv) $W_{curv1}$ is mostly negative; v) $W_{curv2}$ is the largest positive contributor in the downstream region and is generally negated by; vi) $W_{beta}$, which is the largest negative contributor, essentially cancelling $W_{curv2}$ everywhere. In \figref{fig:Sum}, we present the sums of the guiding center drifts, $W_0$, and the Eulerian drifts, $W_1$ in the top and middle panels. For comparison, we again plot $\V{j}_e\cdot\V{E}$. $W_0$ is entirely dominated by $W_{curv0}$ and does not follow the actual $\V{j}_e\cdot\V{E}$ energization; whereas, $W_1$ is a nearly perfect match to the electron work.

\section{Discussion and Conclusions}\label{sec:conclusions}
The guiding-center description of electron energization, \eqr{eq:total_kinetic}, clearly shows that curvature drift, $W_{curv0}$, is the dominant energization mechanism in the reconnection simulation, both globally and locally, away from the x-point. However, neither globally nor locally does the sum of the energization mechanisms agree well with $\V{j}_e\cdot\V{E}$, suggesting that the Lagrangian description, unsurprisingly, does not work well in an Eulerian simulation. The disagreement can be partially ameliorated by including additional guiding-center drifts in \eqr{eq:single_part}, e.g., polarization, and properly transforming from guiding-center\new{s} to the fluid frame.

On the other hand, the Eulerian energization perspective, \eqr{eq:fluidEnergization}, is a rigorously complete description of all bulk-drift motion. This perspective suggests that the diamagnetic drift, $W_{diam}$, dominates globally with large local spikes of agyrotropic drift, $W_{agyro}$, energization. We note that both of these drifts are unique to the Eulerian perspective and have no direct equivalent in a guiding-center model, even when summed over ensembles of guiding centers. Despite being rigorously complete, the nomenclature and grouping of drift terms in \eqr{eq:fluidEnergization} is somewhat ambiguous, as demonstrated by the additional formulations in Eqs. \ref{eq:fluidEnergization2} and \ref{eq:fluidEnergization3}. We also note that the diamagnetic drift can be separated into $\nabla B$ and magnetization drifts, c.f., Appendix F of \citet{Juno:2021}. 

In the Eulerian perspective, the curvature drift, $W_{curv1} = W_{curv2} + W_{beta-curv}$, manifestly depends on the pressure anisotropy, differing significantly from the guiding-center description, which depends on only $p_\parallel$. The dependence on the pressure anisotropy leads to significant local and global cancellation, with $W_{curv2} = p_\parallel \V{u}_E \cdot \B{\kappa}$ and $W_{beta-curv} = -p_\perp \V{u}_E \cdot \B{\kappa}$ nearly cancelling each other everywhere in the domain. The strong negative correlation between the two components suggests that they should be combined as $W_{curv1}$ rather than separated. Although $W_{curv1}$ is generally negative in the simulation, clearly indicating that the curvature drift leads to a net loss of energy, that does not mean that high energy particles are not being produced in the Eulerian picture. Since the Eulerian perspective provides information about bulk energization, it is possible for $W_{curv2}$ to produce an electron power-law tail leading to a bulk drift while $W_{beta-curv}$ produces a greater loss of bulk flow energy from the core of the distribution, resulting in an overall loss of energy due to the curvature drift. Further, we note that in the pressure isotropic limit, the fluid curvature drift mostly vanishes---the component of the curvature drift proportional to $u_\parallel^2$ that formally appears in the polarization drift remains in the pressure isotropic limit. Indeed, this cancellation of curvature-drift terms in a pressure isotropic fluid is even noted in Goldston and Rutherford’s textbook \cite{Goldston:1995}.

The guiding-center formulation can be manipulated to more closely resemble the Eulerian description, including the appearance of the $W_{beta-curv} = -p_\perp \V{u}_E \cdot \B{\kappa}$ term in \eqr{eq:total_kinetic2}. However, this manipulation is not well motivated physically and obfuscates the relevant guiding-center physics by re-expressing betatron acceleration as a series of other terms. The re-expression of the betatron term does, however, serve to highlight the problematic nature of attempting to apply a guiding-center description to an Eulerian system, whether that Eulerian system is a simulation or the solar wind. Because Eqs. \ref{eq:total_kinetic} and \ref{eq:total_kinetic2} are derived by summing over ensembles of guiding centers, fluid moments like pressure and bulk flow arise, which provides a false impression that these guiding-center descriptions can be connected directly to fluid-like bulk evolution of the plasma (Eqs. \ref{eq:total_kinetic} and \ref{eq:total_kinetic2}), but that is not simply the case. Terms like the diamagnetic drift and agyrotropic drift will always be absent \new{from} any formal derivation based on guiding-center motion since guiding-center (or even particle) descriptions cannot account for bulk effects. Further, the guiding-center energization equation, \eqr{eq:single_part}, includes only the motion (drift) of the guiding center plus an induction term ($\mu \partial B / \partial t$) that is due to the electric field acting on the gyroperiod integrated motion of an electron, which is not the complete motion of the particle. The motion of the particle in the \new{reference frame in which the guiding-center is at rest} is not included in \eqr{eq:single_part}, but to connect to an Eulerian \new{perspective}, this motion must be \new{included}. When this motion is included \cite{Goldston:1995}, one arrives at the Eulerian description described herein for terms like the curvature drift. 

So, where does this leave us? Are the prior PIC results indicating that curvature drift leads to significant particle energization incorrect? The short answer is no, because the particle guiding centers are indeed energized (accelerated) by $W_{curv0}$, meaning that where that term is positive and large, particle guiding centers are gaining significant energy. However, that is not the full story, because the analysis typically leading to that conclusion is incomplete and typically performed in an Eulerian \new{framework} by integrating over regions of particles---this fact applies to both PIC simulations and spacecraft data. In the Eulerian \new{perspective}, one must also include additional terms arising from the full particle motion, not just the guiding-center drift plus gyroperiod integrated motion as included in the guiding-center energization equation. When doing so, the full Eulerian curvature drift, $W_{curv1}$, may indicate a very different scenario for the electrons. In the case of the simulation examined herein, $W_{curv1}$ is relatively small and negative, whereas $W_{curv0}$ is large, positive, and the dominant contribution to electron energization. Fundamentally, this contradiction arises because the Eulerian picture describes the generation of bulk flows rather than guiding-center acceleration described by \eqr{eq:single_part}. The confusion follows from attempting to directly relate guiding-center acceleration described by \eqr{eq:single_part} to bulk quantities described by the guiding-center ensemble integrated \eqr{eq:total_kinetic}. The guiding-center ensemble equations do not include the full particle motion, nor do they account for bulk plasma properties that lead to bulk drifts, e.g., diamagnetic and agyrotropic effects. Therefore, the guiding-center ensemble energization \eqr{eq:total_kinetic} can be useful for highlighting regions of the plasma in which electrons are being accelerated by the curvature drift, potentially leading to highly energetic electrons via Fermi acceleration \cite{Drake:2006,Dahlin:2015,Dahlin:2017}; however, the guiding-center ensemble energization does not necessarily indicate regions in which electron bulk flows are large nor where electron heating (thermal broadening of the electron distribution) may be occurring. On the other hand, the Eulerian framework (\eqr{eq:fluidEnergization}) only indicates where electrons are being energized in a bulk sense (bulk flows and heating), since in the Eulerian picture, there is no conception of single-particle or guiding-center motion.

In summary, the guiding-center approach can provide insights into high-energy particles, while the Eulerian approach gives a more complete perspective on the bulk energization and heating of the plasma. The Eulerian picture does describe the production of high energy particles, but only in a bulk sense as power-law tails of the distribution function, which in terms of moments, also lead to bulk flows. The guiding-center description of heating is incomplete, while the Eulerian approach provides no information about individual particle energization to high energies. We thus encourage a greater awareness of the desired goals when applying these diagnostics to the analysis of spacecraft data and simulations so that the proper diagnostic and perspective is applied. Finally, we note that field-particle correlations, by highlighting the different regions of velocity space in which particles are energized, may help to bridge this gap between the Lagrangian and Eulerian pictures \cite{Juno:2021}.

\begin{acknowledgments}
The authors thank Hui Li for helpful discussions.  This research is part of the Frontera computing project at the Texas Advanced Computing Center. Frontera is made possible by National Science Foundation (NSF) award OAC-1818253. JMT was supported by the NSF under grant number AGS-1842638. GGH was supported by the NSF grant AGS-1842561. J. Juno was supported by the U.S. Department of Energy under Contract No. DE-AC02-09CH1146 via an LDRD grant. The development of Gkeyll was partly funded by the NSF-CSSI program, Award Number 2209471.

\end{acknowledgments}

\section*{Data Availability Statement}

\gke\ is open source and can be installed by following the instructions on the \gke\ website (http://gkeyll. readthedocs.io). The input file for the \gke\ simulation presented here is available in the following GitHub repository, https://github.com/ammarhakim/gkyl-paper-inp.

\appendix

\section{Alternative Guiding Center Formulation}\label{sec:appendix}
We begin by noting that the betatron, $\mu \partial B / \partial t$, term appearing in \eqr{eq:single_part} corresponds to induction and is due to the electric field acting on an electron during over its gyroperiod, i.e., it is due to particle motion beyond simple guiding-center motion. We can take advantage of this fact and note that
\begin{eqnarray}
    \frac{\partial B}{\partial t} &=& 
    \mvec{b}\cdot \frac{\partial \mvec{B}}{\partial t} = 
    -\mvec{b}\cdot \nabla \times \mvec{E} = 
    -\mvec{b}\cdot \nabla \times \mvec{E}_\parallel + \mvec{b}\cdot \nabla \times (\mvec{u}_E \times \mvec{B})  \nonumber \\ 
    &=& -\mvec{E}_\parallel \cdot \nabla \times \mvec{b} -B \nabla \cdot \mvec{u}_E - \mvec{u}_E \cdot \nabla B - B \mvec{u}_E\cdot \nabla_\parallel \mvec{b},
\end{eqnarray}
where we have used the fact that in the guiding-center limit, $\mvec{E} = \mvec{E}_\parallel - \mvec{u}_E \times \mvec{B}$. Therefore, \eqr{eq:total_kinetic} can be expressed as
\begin{eqnarray}
    \frac{d E_e}{dt} = &j_\parallel E_\parallel (1 - \beta_\perp / 2)
    - p_\perp \nabla \cdot \mvec{u}_E - p_\perp \mvec{u}_E \cdot \B{\kappa}
    + \nonumber \\ 
    &(p_\parallel + m n u_\parallel^2) \mvec{u}_E \cdot \B{\kappa},
\end{eqnarray}
$\beta_\perp = 2\mu_0 p_\perp / B^2$. With this substitution, it is apparent that a curvature-related term, $W_{beta-curv} = -p_\perp \mvec{u}_E \cdot \B{\kappa}$, was hidden within the betatron term. We also note that  $-p_\perp \nabla \cdot \mvec{u}_E$ is a compressional term similar in form to that studied by \citet{Li:2018}. However, we further note that $\nabla \cdot (p_\perp \mvec{u}_E) = p_\perp \nabla \cdot \mvec{u}_E + \mvec{u}_E \cdot \nabla p_\perp$, which integrated over all space implies $-\int_\mvec{R} p_\perp \nabla \cdot \mvec{u}_E d\mvec{r} = \int_\mvec{R} \mvec{u}_E \cdot \nabla p_\perp d\mvec{r} = \int_\mvec{R} W_{diam} d\mvec{r}$, assuming there is no net flux of thermal energy into or out of the domain. Therefore, in a spatially integrated sense, as often presented in guiding-center analyses, this compressional betatron-related term is related to the Eulerian diamagnetic drift energization. Thus, the electron guiding-center energization can be expressed approximately as
\begin{eqnarray}\label{eq:total_kinetic2}
    \frac{d E_e}{dt} &=  W_{par} + W_{comp} + W_{beta-curv} + W_{curv0} \nonumber \\    
    &\simeq W_{par} + W_{diam} + W_{beta-curv} + W_{curv0}.
\end{eqnarray}

%

\end{document}